\documentclass[letterpaper]{article}
\usepackage{aaai21}  
\usepackage{times}  
\usepackage{helvet} 
\usepackage{courier}  
\usepackage[hyphens]{url}  
\usepackage{graphicx} 
\usepackage{booktabs}
\usepackage{natbib}  
\usepackage{caption} 
\usepackage{multicol}
\usepackage{subcaption}

\begin{document}

\title{NELA-Local: A Dataset of U.S. Local News Articles for the Study of County-level News Ecosystems}
\author {
    Benjamin D. Horne,\textsuperscript{\rm 1}
    Maur\'{i}cio Gruppi,\textsuperscript{\rm 2}
    Kenneth Joseph,\textsuperscript{\rm 3}
    Jon Green,\textsuperscript{\rm 4}
    John P. Wihbey,\textsuperscript{\rm 5} and \\
    Sibel Adal\i  \textsuperscript{\rm 2}\\
}
\affiliations {
    \textsuperscript{\rm 1} School of Information Sciences, University of Tennessee Knoxville, Knoxville, TN, USA\\
    \textsuperscript{\rm 2} Computer Science, Rensselaer Polytechnic Institute, Troy, NY USA \\
    \textsuperscript{\rm 3} Computer Science and Engineering, University at Buffalo, Buffalo, NY USA\\
    \textsuperscript{\rm 4} Network Science Institute, Northeastern University, Boston, MA USA\\
    \textsuperscript{\rm 5} School of Journalism and Media Innovation, Northeastern University, Boston, MA USA\\
    bhorne6@utk.edu, gouvem@rpi.edu, kjoseph@buffalo.edu, \{jo.green, j.wihbey\}@northeastern.edu, adalis@rpi.edu\\
}


\maketitle

\begin{abstract}
In this paper, we present a dataset of over 1.4M online news articles from 313 local U.S. news outlets published over 20 months (between April 4th, 2020 and December 31st, 2021). These outlets cover a geographically diverse set of communities across the United States. In order to estimate characteristics of the local audience, included with this news article data is a wide range of county-level metadata, including demographics, 2020 Presidential Election vote shares, and community resilience estimates from the U.S. Census Bureau. The \texttt{NELA-Local} dataset can be found at: \url{https://dataverse.harvard.edu/dataset.xhtml?persistentId=doi:10.7910/DVN/GFE66K}.
\end{abstract}


\section{Introduction}

Local news, news that primarily serves a specific geographic region \cite{abernathy2018expanding}, is a fundamental component of the larger news ecosystem. Local news organizations can boost civic engagement, investigate wrong-doing, and inform decision making during crisis events, all at a community-specific level that national news outlets cannot fulfil \cite{hendrickson2019local, gollustLocalTelevisionNews2017, chauhan2017providing}. This importance has been recently highlighted by the COVID-19 pandemic, in which, despite being a global event, local conditions are needed in decision making for both community members and public health experts \cite{gollustTelevisionNewsCoverage2019, branswellWhenTownsLose2018}. 

Most existing work has focused on newsrooms adapting to the increased role of digital news consumption \cite{jenkins2021changing}, and the alarming number of regions that no longer have local news outlets, often termed \textit{news deserts} \cite{abernathy2018expanding}. There is also a rich area of work focused on ownership. According to \citet{abernathy2016rise}, for example, ``since 2004 more than a third of the [U.S.] newspapers have changed ownership.'' Many of the outlets serving small to mid-sized communities have been bought and are operated by investment groups, often eroding the connection between the newspapers and local issues. With this shift in ownership comes diminished staffing and a diminished investment in operations in the name of profitability \cite{abernathy2016rise}. Even if these cuts do not impact the sheer volume of news produced by local outlets, they can impact how labor-intensive those stories are, which in turn can reduce community knowledge of local events and the amount of substantive content produced\footnote{\url{https://www.fcc.gov/general/information-needs-communities}}. In other words, local news produced may not actually meet the information needs of the local community.

Perhaps just as important, ownership values can not only change the volume and quality of news produced, but also change the news coverage itself. For example, there is evidence that ownership change can lead to local news coverage increasing around national politics, rather than the local area being served \cite{martin2019local}. Another example of this phenomenon has been termed the \textit{Sinclair Effect} after the wide ownership of local cable news by the Sinclair Broadcast Group. One study found that outlets owned by Sinclair ``produced more stories with dramatic elements, commentary, and partisan sources''  \cite{hedding2019sinclair}.

All of these studies demonstrate that local news is worthy, even crucial, for researchers to study. However, despite local media's known and well documented importance, local news article data is rare, particularly at a large scale over time. Hence, there are very few large scale studies of local news environments. For example, many of the current studies utilize journalist interviews, small, study specific collections that are not made publicly available, or large data about ownership that does not capture article content. While these data sources are certainly useful, many research questions cannot be sufficiently explored using them. For example: How do local outlets differ in the coverage of major events? What sub-topics of coverage are emphasized? How much content is shared across local outlets? What proportions of coverage are dedicated to local issues, rather than national events? To explore these types of large-scale, content-based questions, researchers need stable collections of news article data.

In this paper, we present the \texttt{NELA-Local} dataset, to fill this gap. The \texttt{NELA-Local} dataset contains over 1.4 million online articles from 313 local news outlets in the United States. Uniquely, this data contains nearly every article published by these 313 outlets between April 4th, 2020 and December 31st, 2021. Notably, this timeline covers several historical events, in which local coverage of can be studied, including the 2020 U.S. Presidential Election, the January 6th U.S. capitol riots, and the COVID-19 pandemic. Furthermore, to aid studies using this dataset, we have mapped several open county-level datasets to each outlet, allowing researchers to estimate characteristics about each outlet's local audience. This county-level metadata includes demographics, political leanings, and community resilience estimates. Our hope is that this data can not only fuel novel, large-scale research on local news environments, but also benefit current lines of local media research.

\section{Collection and Mapping Methods}
There are two major parts of the \texttt{NELA-Local} dataset: 1. local news article data and 2. county-level metadata mapped to each outlet based on the county in which the outlet is headquartered. Below we describe the collection and mapping process for each part.

\subsection{Local News Article Data}
The primary, and most unique, part of our dataset is the local news article data. To collect the article data, we take the approach used in \cite{norregaard2019nela}. Namely, we utilize RSS feeds from news websites to collect article URLs, and then follow those URLs to scrape full-text articles. The advantage of this method is that since data is collected as articles are published, nearly all articles published from each outlet can be collected. The challenge of this method is that it requires a predefined list of RSS feed URLs, which can be difficult to obtain without manual effort. 

To this end, we start with a publicly available list of U.S. local news outlets\footnote{\url{www.50states.com/news/}}. This list comes from a social studies classroom resources website (www.50states.com). From our understanding, this list has been regularly expanded between 1996 and until at least 2018.

Using this list, we search for those that have active RSS feeds. This search is done using a mix of automatic and manual methods. First, we write a script that iterates over the website URLs and checks if common RSS feed paths exist within the website. For example, \texttt{www.examplenewswebsite.com/rss} or \texttt{www.examplenewswebsite.com/feed}. Then the script checks for paths on commonly used web feed management services like Feed Burner. Second, after the script is complete, two authors manually checked the found RSS feeds to ensure they appeared to be up-to-date and checked the websites without found RSS feeds for unusual RSS feed paths or services.

From this search, we found that 568 of the 3,300 outlets on the site had active RSS feeds. Next, one of the authors, who is an expert in local news, manually checked these outlets to ensure they were legitimate local news outlets. This manual filter removed 12 more outlets from the 568. Outlets that were removed that did not match the definition
of local news \cite{abernathy2016rise}. Specifically, that the website was a dedicated news site that primarily targets a geographically local audience. Hence, the outlets removed were either specifically for university students, explicitly national outlets, or were not news-oriented.

Using this filtered list of 556 outlets, we scrape each outlets' RSS feed twice a day, everyday between April 4th, 2020 and December 31st, 2021. Importantly, as the RSS feeds are crawled, we scrape full article text by following the URLs to the web pages that have the full article text, rather than scraping snippets of articles that may appear in the RSS feeds. This process is shown in Figure \ref{fig:flow}.

We then perform several secondary robustness checks on the scraped data from the 556 outlets, reducing the final set of outlets to 313. First, we removed any outlets that had less than 50 articles during the time frame, indicating an outlet that publishes infrequently or does not maintain their RSS feed regularly. Second, we removed outlets that publish only in Spanish. Last, we removed sources that were later found to maintain multiple RSS feeds per news topic, but only one of the feeds were scraped (for example, one feed for Covid-19 news, another for general news). In some cases, these split-off RSS feeds were created after our data collection started. Importantly, this last step ensures that the data is of high enough quality to do analysis of topical coverage over time without artificially over-representing a topic (see Section \ref{sec:usecases} for use case discussion). 

The final dataset has 313 outlets and 1,445,509 articles.


\subsection{County-level Metadata}\label{sec:countymeta}
The second part of our dataset is county-level metadata. Specifically, we map each of the final 313 outlets in our dataset to the county in which it is headquartered, allowing us to utilize several open datasets. We obtain the Federal Information Processing Standard code (FIPS) for each county, and use the FIPS code to map outlets to the county-level datasets. The goal of this mapping is gain an approximate understanding of each outlets audience, similar to the process done in dataset presented in \cite{abernathy2016rise}. Note, there are fewer counties than there are outlets (255 counties and 313 outlets) as counties can have more than one outlet. Below we describe each of the county-level datasets.

\paragraph{Demographics} 
First, we map demographic data for each outlet using MIT Election Data and Science Lab's Codebook for 2018 Election Analysis Dataset\footnote{\url{https://github.com/MEDSL/2018-elections-unoffical/blob/master/election-context-2018.md}}. Broadly, data we include from this dataset are: population, race, age, education, and economic information, including median household income and unemployment. Each of these data are described in detail in Table \ref{tbl:demo} in the Appendix.

Note, this dataset does not include Alaska. However, since we collected rich, complete news article data from three outlets in Alaska, we choose to keep them in the dataset. Instead of removing the data all together, we place 'Null' values for these counties in the demographics table (see Figure ~\ref{fig:schema}) and `None' values in the demographic columns of the CSV file (see Section \ref{sec:csv}).

\paragraph{Politics} 
Second, we map county-level political leanings to each outlet. To do this, we link each outlet to the vote share of the last three U.S. Presidential elections: Biden vs. Trump in 2020, Clinton vs. Trump in 2016, and Obama vs. Romney in 2012. This data comes from two publicly available datasets. For the 2016 and 2012 election data, we again use MIT Election Data and Science Lab's Codebook for 2018 Election Analysis Dataset\footnote{\url{https://github.com/MEDSL/2018-elections-unoffical/blob/master/election-context-2018.md}}. For the 2020 election data, we use Kieran Healy's 2020 Election Results data\footnote{\url{https://github.com/kjhealy/us_elections_2020_csv}}.

To ensure each counties vote share is comparable and standardized, we store each as the log-odds of a person in the county voting for the Republican candidate in each election. Hence, for example, higher log-odds for Biden vs. Trump in 2020 means that the county leaned towards Trump, or higher log-odds for Obama vs. Romney in 2012 means that the county leaned towards Romney.

Note, just like the demographic data, counties in Alaska are not included in these two datasets. We again place `Null' values for these counties in the politics table (see Figure ~\ref{fig:schema}) and `None' values in the politics columns of the CSV file (see Section \ref{sec:csv}).

Each of these data are described in detail in Table \ref{tbl:poli} in the Appendix.

\paragraph{Risks} 
Third, to provide an assessment of \textit{risks} per outlet audience, we map our data to the U.S. Census Bureau's \textit{Community Resilience Estimates}\footnote{\url{https://www.census.gov/programs-surveys/community-resilience-estimates.html}}. According to the U.S. Census Bureau, these Community Resilience Estimates ``provide an easily understood metric for how at-risk every neighborhood in the United States is to the impacts of disasters, including COVID-19 \cite{}.'' More specifically, the estimates provided fall into three groups: the percentage of individuals had 0 risk factors, 1-2 risk factors, or 3+ risk factors, where a higher number of individuals in lower risk factors indicates a community that is more resilient to disasters. These estimates are determined by examining demographic, socioeconomic, and housing characteristics in the American Community Survey (ACS) microdata \footnote{\url{https://www.census.gov/programs-surveys/community-resilience-estimates/technical-documentation/methodology.html}}. These characteristics included a variety of factors such as: household Income-to-Poverty Ratio (IPR), the number of caregivers per household, Unit-level crowding per household, health insurance coverage, and vehicle access. 

All 313 outlets in our dataset are mapped to the risks data. Each of these data are described in detail in Table \ref{tbl:risks} in the Appendix.

\section{Data Formats}
In order to accommodate the widest audience possible, we provide several data formats. 

\subsection{Normalized SQLite3 Database}
\begin{figure*}[h]
    \centering
    \begin{subfigure}{.45\textwidth}
        \centering
        \includegraphics[width=6cm]{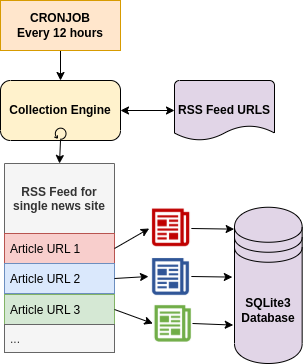}
        \caption{Article data collection flow}
        \label{fig:flow}
    \end{subfigure}
    \begin{subfigure}{.5\textwidth}
        \centering
        \includegraphics[width=9.5cm]{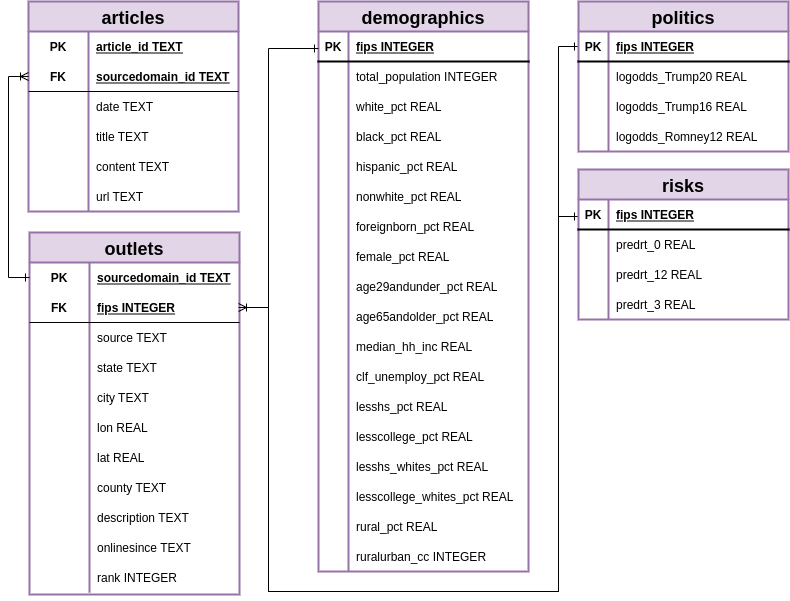}
        \caption{Database schema}
        \label{fig:schema}
    \end{subfigure}
\caption{In \textbf{(a)} we show the flow of our news article collection. The general flow is as follows: 1. Every 12 hours start the collection engine. 2. Collection engine ingest and loops through a predetermined list of RSS feed links. 3. Each RSS feed is parsed for new article URLs, those URLs are followed to scrape the full article text from the outlets website. 4. New articles are stored in the SQLite3 database. In \textbf{(b)} we show the SQLite3 database schema. The main table is the \texttt{articles} table, which is mapped to a table for outlets and their location. Each outlet in the \texttt{outlet} table is then mapped to three different county-level metadata tables: \texttt{demographics}, \texttt{politics}, and \texttt{risks}}.
\end{figure*}

The first format is a normalized, SQLite3 database with five tables: articles, outlets, demographics, politics, and risks. The schema for this database can be found in Figure \ref{fig:schema}. The primary table in the database is the \texttt{articles} table, which includes the title, content, date, and URL for each article. Additionally, the \texttt{articles} table includes a identifier for the outlet of the article (called sourcedomain\_id), which is the foreign key for the \texttt{outlets} table. The \texttt{outlets} table includes data about the location of the outlet, including the FIPS code. The FIPS code in the outlet table maps to all three of the county-level metadata tables: \texttt{demographics}, \texttt{politics}, and \texttt{risks}. 

In addition to the database, we provide a short example Python script to use the database. 

\subsection{CSV}\label{sec:csv}
The second format we provide the dataset in is Comma-Separated Value (CSV) files. Specifically, we provide five CSV files, one for each table in the database: articles, outlets, demographics, politics, and risks. The columns in each CSV file are the same as the columns in each corresponding SQLite3 database table. 

\subsection{FAIR Principles}
The \texttt{NELA-Local} dataset follows FAIR principles\footnote{\url{https://www.force11.org/group/fairgroup/fairprinciples}}. Namely, the data \textit{Findable}, as it is persistently stored on Harvard Dataverse and is described with rich metadata. The data is \textit{Accessible} and \textit{Interoperable}, as it is retrievable through Harvard Dataverse's GUI and is stored in two widely-used, standard formats (SQLite3 and CSV), including sample Python scripts for extraction. Given the formats, the data can be parsed by both machine and human annotators. Additionally, this paper describes metadata and points users to other datasets that can complement and augment this dataset (see Section \ref{sec:related}). Lastly, the data is \textit{Re-usable}, as it is open for free use, ready to use out-of-the-box in a wide range of local news studies, and contains original article URLs and metadata links/documentation to maintain provenance.

\section{Data Quality and Descriptive Statistics}
To demonstrate the quality of our dataset, we provide several sets of descriptive statistics. There are two core traits that demonstrate the quality of this dataset: 1. The data comes from a geographically and demographically diverse set of counties. 2. The data contains nearly every article published by each outlet over time.

\begin{figure}[h]
    \centering
    \includegraphics[width=8cm]{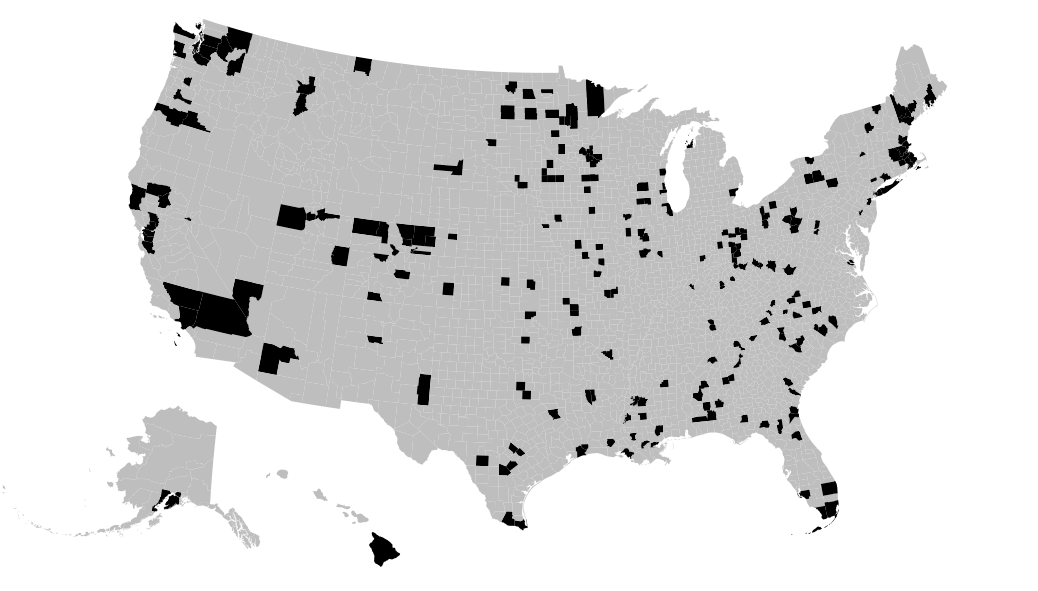}
    \caption{Map of counties in dataset, where black indicates an area where our dataset has at least 1 outlet.}
    \label{fig:map}
\end{figure}

\begin{figure*}[h]
\centering
\begin{subfigure}{.33\textwidth}
  \centering
  \includegraphics[width=5.5cm]{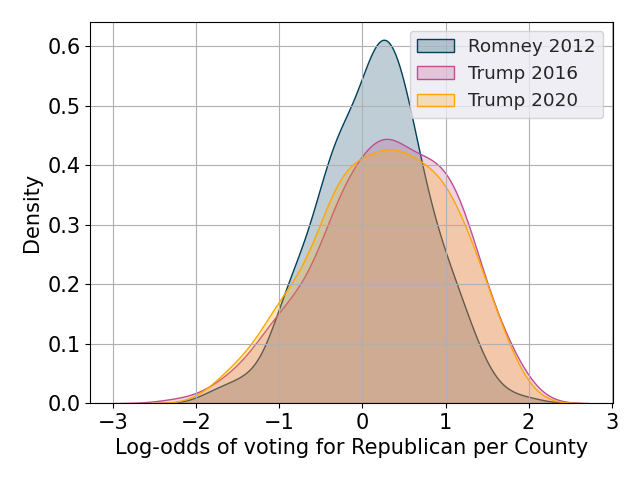}
  \caption{Log-odds vote per county}
  \label{fig:vote}
\end{subfigure}%
\begin{subfigure}{.33\textwidth}
  \centering
  \includegraphics[width=5.5cm]{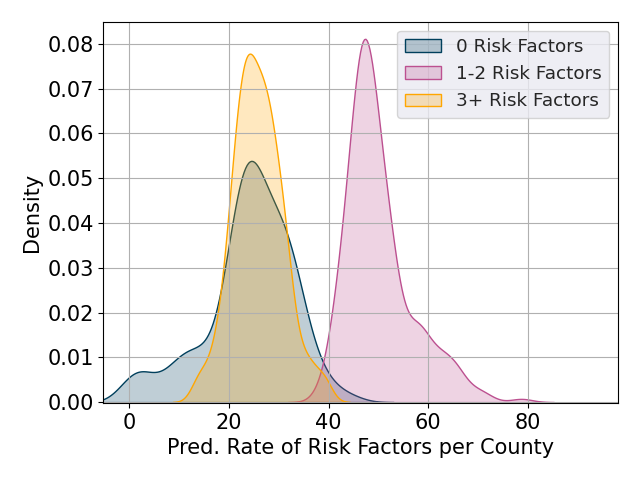}
  \caption{Pred. rate of risk factors per county.}
  \label{fig:risks}
\end{subfigure}
\begin{subfigure}{.33\textwidth}
  \centering
  \includegraphics[width=5.5cm]{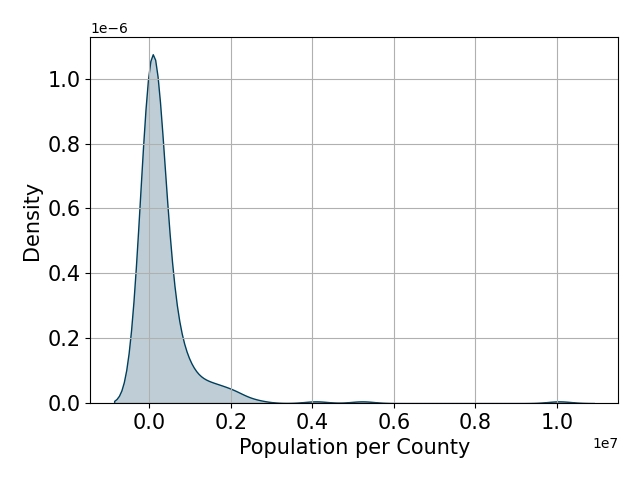}
  \caption{Population per county}
  \label{fig:pop}
\end{subfigure}
\begin{subfigure}{.33\textwidth}
  \centering
  \includegraphics[width=5.5cm]{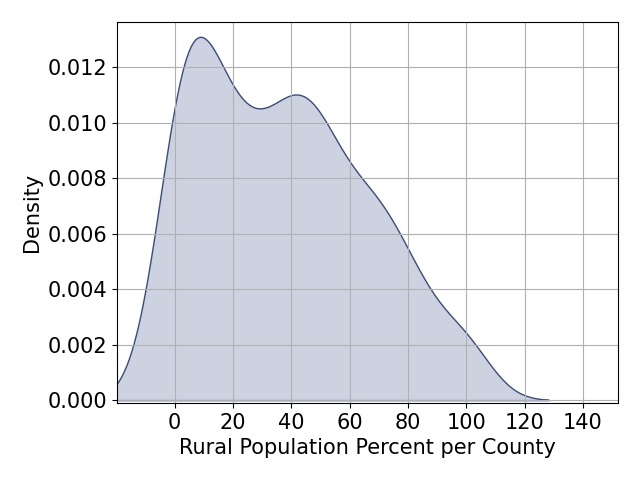}
  \caption{Rural population percent per county}
  \label{fig:rural}
\end{subfigure}%
\begin{subfigure}{.33\textwidth}
  \centering
  \includegraphics[width=5.5cm]{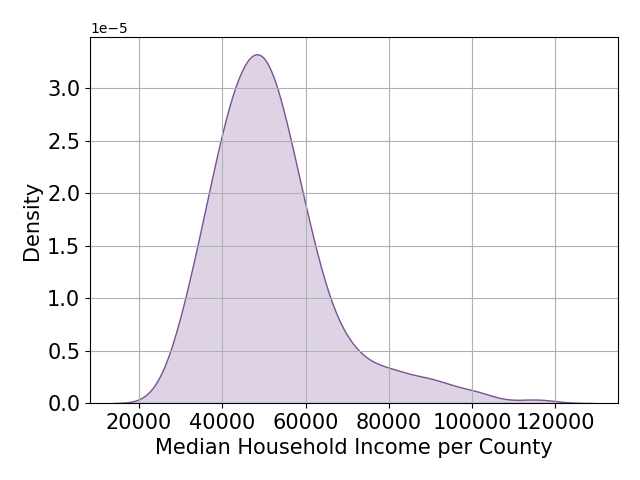}
  \caption{Median household income per county}
  \label{fig:hhinc}
\end{subfigure}
\begin{subfigure}{.33\textwidth}
  \centering
  \includegraphics[width=5.5cm]{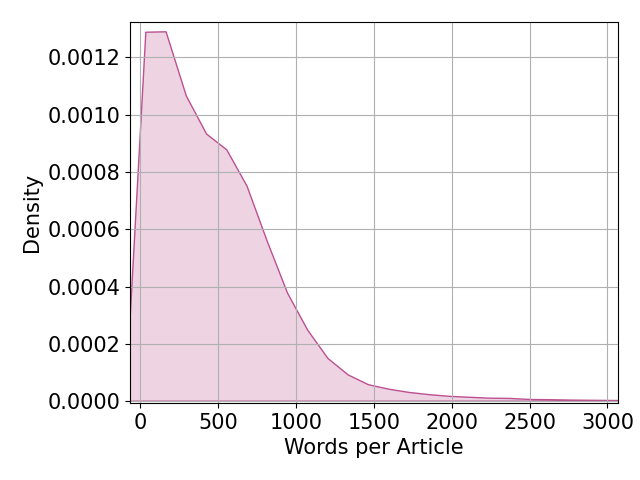}
  \caption{Words per article}
  \label{fig:words}
\end{subfigure}
\begin{subfigure}{.33\textwidth}
  \centering
  \includegraphics[width=5.5cm]{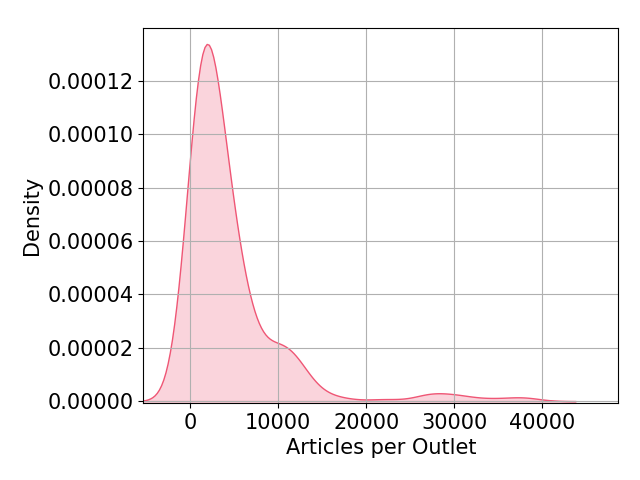}
  \caption{Articles per outlet}
  \label{fig:apero}
\end{subfigure}%
\begin{subfigure}{.33\textwidth}
  \centering
  \includegraphics[width=5.5cm]{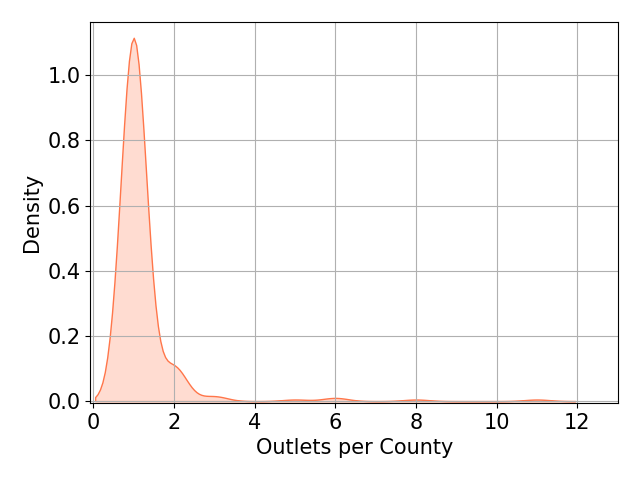}
  \caption{Outlets per county}
  \label{fig:operc}
\end{subfigure}
\begin{subfigure}{.33\textwidth}
  \centering
  \includegraphics[width=5.5cm]{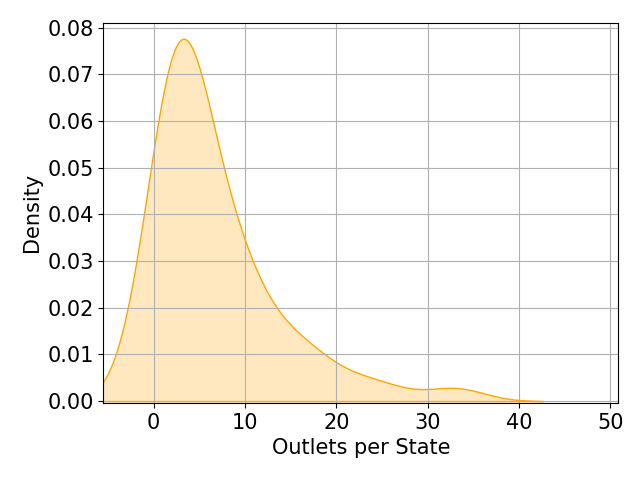}
  \caption{Outlets per state}
  \label{fig:opers}
\end{subfigure}
\caption{Kernel Density Estimate plots for \textbf{(a)} Log-odds of a person in a county in which a outlet is situated voting for the Republican Presidential Candidate in 2012, 2016, and 2020, \textbf{(b)} Estimated percentage of the population with 0, 1 to 2, or 3+ disaster/health related risks per county in the dataset, \textbf{(c)} Population of counties in the dataset, \textbf{(d)} Percent of population that is rural per county in the dataset, \textbf{(e)} Median household income per county in the dataset, \textbf{(f)} Words per article in the dataset, \textbf{(g)} Articles per outlet in the dataset, \textbf{(h)} Outlets per county in the dataset, and \textbf{(i)} Outlets per state in the dataset. }
\label{fig:meta}
\end{figure*}

\paragraph{Geographical and demographic diversity} In Figure \ref{fig:map}, we show a map of the counties in which our dataset contains at least 1 local news outlet. In total, our dataset contains outlets in 255 counties from 46 states (States not covered are: Delaware, Idaho, Maryland, and Wyoming). On average, the dataset has 1.23 outlets per county and 6.80 outlets per state (see KDE plots in Figure \ref{fig:operc} and \ref{fig:opers}). The max number of outlets in a single county is 11 (Middlesex County, MA) and the max number of outlets in a single state is 33 (Massachusetts). Two notes about the skew towards Massachusetts: First, 20 of the outlets in Massachusetts are all run by the same parent company, WickedLocal, and could be combined or removed depending on the analysis being done. However, these 20 sites do produce different content. Second, Middlesex County is the most populous county in both Massachusetts and New England\footnote{\url{https://data.census.gov/cedsci/profile?g=0500000US25017}}. 

In addition to geography, we see diversity in various demographic factors. On average, the populations of counties in our dataset are 37.6\% rural, with the minimum percent rural being 0\% and the maximum being 100\% (See KDE plot in Figure \ref{fig:rural}). To add context, an example of a county with 0\% rural population in our dataset is Denver County, Colorado and a county with 100\% is Lewis County, Kentucky. On average, counties contain a population of 355,998.82, with a minimum population of 4347 and maximum population of 10,057,155. Furthermore, the average median household income per county is \$52,077.22, with a minimum of \$28,136 and maximum of \$115,224 (see KDE plot in Figure \ref{fig:hhinc}). Again to add context, according to the Census Bureau, the U.S. median household income in 2018 was \$63,179\footnote{\url{https://www.census.gov/library/stories/2019/09/us-median-household-income-not-significantly-different-from-2017.html}}. 

Similarly, the political leanings of counties in our dataset are diverse. Namely, on average, counties had a 0.25 log-odds of voting for Trump in 2020, with a minimum log-odds of -1.75 (left leaning) and a maximum log-odds of 1.83 (right leaning). We see very similar vote shares in 2016 and 2012. On average counties had a 0.29 log-odds of voting for Trump in 2016 and a 0.12 log-odds of voting for Romney in 2012. For the full distribution, see KDE plot in Figure \ref{fig:vote}.

Hence, the \texttt{NELA-Local} dataset covers a wide range of outlet audiences, covering a diverse set of rural and urban settings. Overall, the audiences covered lean slightly right politically and have a slightly lower median household income than the U.S. as a whole. 

\begin{figure*}[h]
\centering
\begin{subfigure}{.33\textwidth}
  \centering
  \includegraphics[width=5.5cm]{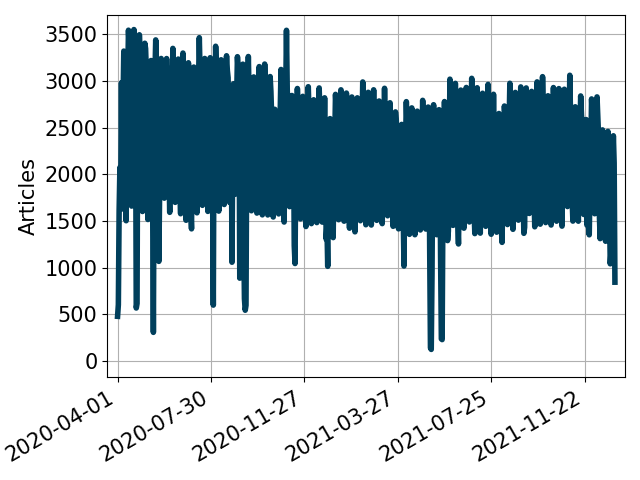}
  \caption{Articles per day}
  \label{fig:day}
\end{subfigure}%
\begin{subfigure}{.33\textwidth}
  \centering
  \includegraphics[width=5.5cm]{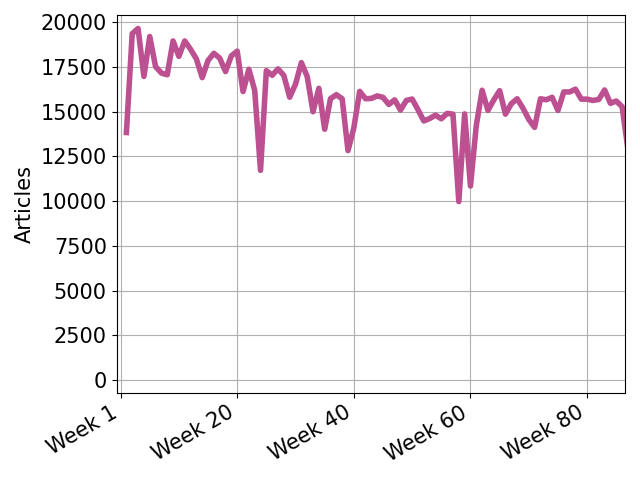}
  \caption{Articles per week}
  \label{fig:week}
\end{subfigure}
\begin{subfigure}{.33\textwidth}
  \centering
  \includegraphics[width=5.5cm]{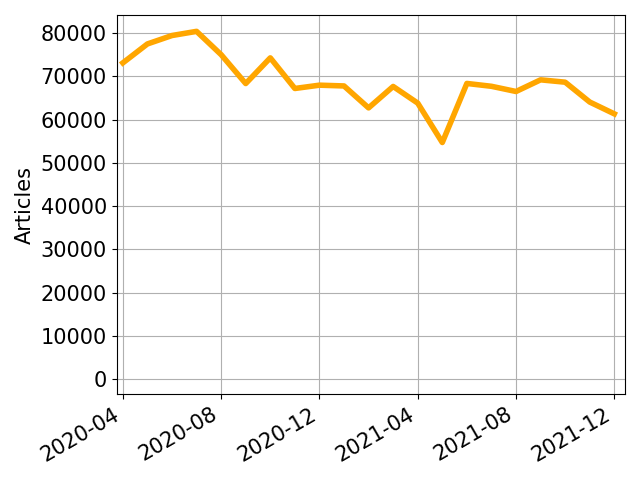}
  \caption{Articles per month}
  \label{fig:month}
\end{subfigure}
\caption{Articles over Time: (a) Per Day, (b) Per Week, (c) Per Month. There is one known data collection outage during May 2021 due to a cyber attack where our collection server was located, however, we were able to recover most of this missed data.}
\label{fig:time}
\end{figure*}

\paragraph{Completeness over time} In Figure \ref{fig:time}, we show the volume of articles across time. Specifically, we show the number of articles per day (a), per week (b), and per month (c). There is one known data collection outage in May 2021 due to a cyber attack where our collection server was located\footnote{\url{https://www.news10.com/top-stories/rpi-restores-network-access-following-cyber-attack/}}. However, we were able to recover most of the data missing during that time, but not all, hence the significant dip during that time. To the best of our knowledge, the rest of the timeline follows the ebbs and flows of publishing patterns per outlet. Some outlets publish daily, while others publish weekly or bi-weekly. 

\section{Use Cases}\label{sec:usecases}
Below we provide a set of three studies that we believe would benefit from our dataset. While these studies emphasize the large-scale temporal nature of the dataset, it is also true that the data can be used for smaller, in-depth qualitative  studies (or mixed-methods approaches) as well. Nor, of course, is this discussion of use cases exhaustive. 

\subsection{Local Media Coverage of Disasters and Events}
One critical use of the presented dataset could be to examine local media coverage of disasters and events. A long-standing literature on agenda setting and framing demonstrates that both the presentation and frequency of coverage of such events influences the importance that the public assigns to them \cite{entman1993framing, scheufele2007framing}. Media coverage of disasters may also shed light on some aspects of the event, while leaving other details out of the coverage, therefore influencing what audiences believe about the event \cite{harbert2010agenda}. Disasters are inherently localized, making local coverage of those disasters important in community sense-making, particularly during the uncertainty of a crisis event \cite{gollustTelevisionNewsCoverage2019, krafft2017centralized}. 

As an example of this use of these data,  \citet{joseph2021local} use an early subset of the \texttt{NELA-Local} dataset to examine the relationship between local news coverage of COVID-19 and local conditions. By mapping a subset of the presented dataset to county and state level COVID-19 case counts, deaths, and politics, the authors are able to provide new insights into factors associated with the degree of local COVID-19 coverage overtime and demonstrate how pandemic-related subtopics vary across local areas. 

Given the major national and localized events that occurred in the U.S. during the time frame of our dataset (e.g. 2020 Presidential Election, U.S. capitol riots, local elections), the county-level metadata provided,  and the fact that collection of these data continue over the present, similar studies on the relationship among coverage and local audience can be done ``out-of-the-box.''

\subsection{Measuring the Nationalization of Local News}
By pairing this dataset with national news datasets from similar timeframes (see discussion of Media Cloud and NELA-GT in Section \ref{sec:related}), one can measure the similarity in coverage between national outlets and local outlets. This type of study could be done quantitatively, using text analysis techniques such as those in \cite{starbird2018ecosystem} and \cite{horne2019different}. Or it could be done qualitatively by extracting subsets of stories from each dataset for thematic analysis. While current literature has provided evidence of local media increasingly reporting on national events, particularly in politics, demonstrating this trend over time, topic, and outlets has not been explored.

Similarly, studies examining the impact of local media ownership on coverage can be done by mapping the \texttt{NELA-Local} dataset to ownership information (see discussion of the UNC database in Section \ref{sec:related}).  Finally, along this line, the extent to which different subsets of local news media mirror each other, and the potential causes of this (e.g. county-level demographics) could also be explored with the datasets presented here.

\subsection{Characterizing Hybrid Local Media}
This dataset can add novel contributions to the literature on hybrid media systems \cite{chadwick2017hybrid}, which are systems where old and new media logics, such as traditional news media and social media, are mixed together. Studies on this topic have examined the phenomenon of Twitter content being used in news articles, but these studies have been on small data sets of national news \cite{broersma2013twitter, oschatz2021twitter}, not local news. Because the \texttt{NELA-Local} dataset contains the original URLs to all articles, embedded social media data can be scraped using a similar method to that used in \cite{gruppi2021nela}. If social media content is quoted or used as the source of the article, this will be captured in the article text already in the dataset.

\section{Related Datasets}\label{sec:related}
There are several related, but notably different, datasets to the \texttt{NELA-Local} dataset.

\textbf{Media Cloud} is an open source platform that is used for ``collect[ing] data for studying the media ecosystem on the
open web'' \cite{roberts2021media}. The platform includes several web-based tools that operate on a stored set of media data. The focus of Media Cloud's stored data is distinctly different than \texttt{NELA-Local}. Namely, Media Cloud is focused on capturing global news coverage using a ``combination of automated search ... [and] identified lists of influential sources.\footnote{\url{https://sources.mediacloud.org/#/home}}'' Whereas, our dataset is focused on U.S. local news outlets that serve distinct geographical regions. Hence, Media Cloud contains some overlap in outlets with our dataset, specifically those local outlets headquartered in large population centers, but does not contain outlets located in small, rural population areas. 

\textbf{LexisNexis} is a commercial news database that has been widely used in academic studies \cite{deacon2007yesterday, weaver2008finding}. It being expensive and proprietary is its main downfall. If a university has a subscription to it, stories can be downloaded through an API or the web interface. Because of lack of documentation and its commercial use, it is difficult to assess outlet overlap between LexisNexis data and our dataset. However, LexisNexis is focused on large, formal news sources, such as the Associated Press, which makes it unlikely to track small rural area news outlets. 

There are event-based collections \cite{wang2016growing}, such as \textbf{GDELT} \cite{leetaru2013gdelt} and \textbf{Event Registry} \cite{rupnik2016news}. While the end-goal of these databases are quite different than our dataset, focused on storing events rather than news articles (news articles are used to find the events), they are occasionally used in news and media studies. In particular, GDELT stores full-text data in some cases and is open for academic use. However, GDELT has received criticism for incomplete documentation and lacking coverage of important U.S. sources \cite{kwak2014understanding, weller2014raining}. Event Registry overlaps in outlet coverage with GDELT, but is now a commercial entity \cite{kwak2016two, roberts2021media}. Both event-based do not cover the U.S. Local outlets contained in \texttt{NELA-Local}.

The \textbf{NELA-GT} datasets are static, full-text news article datasets released on a yearly basis \cite{norregaard2019nela,gruppi2020nela,gruppi2021nela}. The primary goal of the NELA-GT datasets are to provide labeled data for machine learning tasks, therefore they include outlet-level veracity labels. Similar to Media Cloud, the NELA-GT datasets contain a large variety of news outlets around the globe and includes low-veracity, disinformation-peddling outlets. While similar in collection method, \texttt{NELA-Local} is not focused on providing data for veracity tasks, but rather providing data to examine local news environments. Furthermore, NELA-GT and \texttt{NELA-Local} contain mutually exclusive sets of outlets. Depending on the scope of study, we believe data from \texttt{NELA-Local}, NELA-GT, and Media Cloud can complement each other and could be used together (although carefully, as they each primarily serve different purposes). 

All of these related data collections discussed above serve a broader, different purpose than the dataset presented in this paper. Our goal is to aid large-scale study of U.S. local news environments, while many of these data collections serve more general purpose, global studies of news media. \textbf{Yin Leon's Local News Dataset} \cite{leon_yin_2018_1345145}\footnote{\url{https://github.com/yinleon/LocalNewsDataset}} may be the closest related dataset in terms of scope. However, the dataset's focus is on TV stations, while ours is on online news articles. The dataset also covers a different time frame. None of the datasets mentioned, including Yin Leon's, contain county-level metadata like \texttt{NELA-Local}. However, Yin Leon's dataset does contain ownership information.

Another dataset that is close in scope and focus to \texttt{NELA-Local} is the \textbf{UNC Database} \cite{abernathy2016rise}. The UNC Database contains information on about publication frequency, circulation statistics, and ownership of 7,927
newspapers. Like the data presented in this paper, the UNC Database contains county-level metadata such as Rural-Urban
Continuum Codes. Unlike the data presented in this paper, the UNC Database does not contain article text data. The two datasets also cover different timeframes. This is another dataset that we believe could be used in conjunction with \texttt{NELA-Local} depending on the study.

\section{Conclusion}
In this paper, we presented a novel dataset of 1.4 million U.S. local news articles mapped to county demographics, politics, and risks data. We argued that the research community lacks large-scale, reliable local news datasets, particularly those containing full-text content for the analysis of topical coverage. By filling this gap, researchers can better understand what types of information local communities are receiving and what types of information those communities are lacking. Futhermore, we provided an extensive discussion of related datasets, some of which can be used to augment the presented dataset for a variety of studies.

The \texttt{NELA-Local} dataset, sample code, and further documentation can be found at: \url{https://dataverse.harvard.edu/dataset.xhtml?persistentId=doi:10.7910/DVN/GFE66K}.

\begin{small}
\bibliography{scibib}
\end{small}

\appendix
\onecolumn
\section{Data Column Descriptions}
In this appendix, we provide detailed descriptions of each data column in the \texttt{NELA-Local} dataset. Below are tables for each table in the database (articles, outlets, politics, risks, and demographics).


\begin{table*}[h]
\centering
\fontsize{9pt}{9pt}
\selectfont
\begin{tabular}{c|p{11cm}}
\textbf{Column Name} & \textbf{Description}\\
\toprule
article\_id & Unique id for article, made up of date, outlet name, and first 100 characters of the article title\\\midrule
sourcedomain\_id & Unique outlet identifier, made up of outlet name and website domain\\\midrule
date & Date the article was published in format MM/DD/YYYY\\\midrule
title &  Title of article\\\midrule
content &  Full text content from article\\\midrule
url &  URL of the article when the article was scraped\\
\end{tabular}
\caption{\textbf{articles} data description}
\label{tbl:art}
\end{table*}

\begin{table*}[h]
\fontsize{9pt}{9pt}
\selectfont
\centering
\begin{tabular}{c|p{11cm}}
\textbf{Column Name} & \textbf{Description}\\
\toprule
 sourcedomain\_id & Unique identifier, made up of outlet name and website domain. Note, some outlets share websites if they have the same parent company.\\ \midrule
 fips & Federal Information Processing Standards code\\\midrule
 source & Outlet name\\\midrule
 description & Description of outlet from website, if available\\\midrule
 onlinesince & How long the outlet has been online, if available \\\midrule
 rank & Amazon Alexa’s ranking of web domains for the outlet website, if available\\\midrule
 state & Name of state outlet is headquartered in\\\midrule
 city & Name of city outlet is headquartered in\\\midrule
 lon & Longitude of county where outlet is headquartered\\\midrule
 lat & Latitude of county where outlet is headquartered\\\midrule
 county & Name of county outlet is headquartered in \\
\end{tabular}
\caption{\textbf{outlets} data description}
\label{tbl:outlets}
\end{table*}

\begin{table*}[h]
\centering
\fontsize{9pt}{9pt}
\selectfont
\begin{tabular}{c|p{11cm}}
\textbf{Column Name} & \textbf{Description}\\
\toprule
 fips & Federal Information Processing Standards code\\\midrule
 logodds\_Trump20 & Log-odds of a person in the county in which the outlet is situated voting for Trump in 2020\\\midrule
 logodds\_Trump16 & Log-odds of a person in the county in which the outlet is situated voting for Trump in 2016\\\midrule
 logodds\_Romney12 & Log-odds of a person in the county in which the outlet is situated voting for Romney in 2012
\end{tabular}
\caption{\textbf{politics} data description. Note, the vote shares data to form the logodds comes from \url{https://github.com/MEDSL/2018-elections-unoffical/blob/master/election-context-2018.md} and \url{https://github.com/kjhealy/us_elections_2020_csv}.}
\label{tbl:poli}
\end{table*}

\begin{table*}[h]
\centering
\fontsize{9pt}{9pt}
\selectfont
\begin{tabular}{c|p{11cm}}
\textbf{Column Name} & \textbf{Description}\\
\toprule
 fips & Federal Information Processing Standards code\\\midrule
 predrt\_0 & Estimated percentage of the population with 0 disaster/health related risks\\\midrule
 predrt\_12 & Estimated percentage of the population with 1 to 2 disaster/health related risks\\\midrule
 predrt\_3 & Estimated percentage of the population with 3+ disaster/health related risks
\end{tabular}
\caption{\textbf{risks} data description. Note, this data comes from \url{https://www.census.gov/programs-surveys/community-resilience-estimates.html}.}
\label{tbl:risks}
\end{table*}

\begin{table*}[h]
\centering
\fontsize{9pt}{9pt}
\selectfont
\begin{tabular}{c|p{11cm}}
\textbf{Column Name} & \textbf{Description}\\
\toprule
 fips & Federal Information Processing Standards code\\\midrule
 total\_population & Population of county, 2012-2016 ACS 5-Year Estimates \\\midrule
 white\_pct & Non-Hispanic whites as a percentage of total population \\\midrule
 black\_pct &  Non-Hispanic blacks as a percentage of total population\\\midrule
 hispanic\_pct & Hispanics or Latinos as a percentage of total population \\\midrule
 nonwhite\_pct &  Non-whites as a percentage of total population \\\midrule
 foreignborn\_pct &  Foreign-born population as a percentage of total population\\\midrule
 female\_pct & Females as a percentage of total population\\\midrule
 age29andunder\_pct & Population 29 years or under as a percentage of total population \\\midrule
 age65andolder\_pct &  Population 65 years or older as a percentage of total population \\\midrule
 median\_hh\_inc & Median household income in the past 12 months (in 2016 inflation-adjusted dollars). \\\midrule
 clf\_unemploy\_pct & Unemployed population in labor force as a percentage of total population in civilian labor force \\\midrule
 lesshs\_pct & Population with an education of less than a regular high school diploma as a percentage of total population\\\midrule
 lesscollege\_pct & Population with an education of less than a bachelor's degree as a percentage of total population\\\midrule
 lesshs\_whites\_pct & White population with an education of less than a regular high school diploma as a percentage of total population\\\midrule
 lesscollege\_whites\_pct & White population with an education of less than a bachelor's degree as a percentage of total population\\\midrule
 rural\_pct & Rural population as a percentage of total population in 2010\\\midrule
 ruralurban\_cc & Rural-urban continuum code from USDA Economic Research Service in 2013. Note, Table \ref{tbl:cc_codes} contains the continuum code descriptions.
\end{tabular}
\caption{\textbf{demographics} data description. Note this data comes from \url{https://github.com/MEDSL/2018-elections-unoffical/blob/master/election-context-2018.md}}
\label{tbl:demo}
\end{table*}

\begin{table*}[h]
\centering
\fontsize{9pt}{9pt}
\selectfont
\begin{tabular}{c|p{11cm}}
\textbf{Column Name} & \textbf{Description}\\
\toprule
 1 & Counties in metro areas of 1 million population or more\\\midrule
 2 & Counties in metro areas of 250,000 to 1 million population\\\midrule
 3 & Counties in metro areas of fewer than 250,000 population\\\midrule
 4 & Urban population of 20,000 or more, adjacent to a metro area\\\midrule
 5 & Urban population of 20,000 or more, not adjacent to a metro area\\\midrule
 6 & Urban population of 2,500 to 19,999, adjacent to a metro area\\\midrule
 7 & Urban population of 2,500 to 19,999, not adjacent to a metro area\\\midrule
 8 & Completely rural or less than 2,500 urban population, adjacent to a metro area\\
 9 & Completely rural or less than 2,500 urban population, adjacent to a metro area\\
\end{tabular}
\caption{Rural-urban continuum codes from USDA Economic Research Service in 2013 (\url{https://www.ers.usda.gov/data-products/rural-urban-continuum-codes/}) used in \texttt{demographics} table.}
\label{tbl:cc_codes}
\end{table*}

\end{document}